# Controlling spin currents with magnon interference in a canted antiferromagnet

Lutong Sheng [1,2,†], Anna Duvakina [3,†], Hanchen Wang [4,†], Kei Yamamoto [5,6,†], Rundong Yuan [1,7,†], Jinlong Wang [1,2], Peng Chen [8], Wenqing He [8], Kanglin Yu [1,2], Yuelin Zhang [1,2], Jilei Chen [9,2], Junfeng Hu [2], Song Liu [2,9], Xiufeng Han [8], Dapeng Yu [2,9], Jean-Philippe Ansermet [10,9], Sadamichi Maekawa [6,11,5], Dirk Grundler [3,12,*] & Haiming Yu [1,2,*]

[1]*Fert Beijing Institute, MIIT Key Laboratory of Spintronics, School of Integrated Circuit Science and Engineering, Beihang University, Beijing 100191, China.*

[2]*International Quantum Academy, Shenzhen 518048, China.*

[3]*Laboratory of Nanoscale Magnetic Materials and Magnonics, Institute of Materials (IMX), École Polytechnique Fédérale de Lausanne (EPFL), Lausanne, Switzerland.*

[4]*Department of Materials, ETH Zurich, Zurich 8093, Switzerland.*

[5]*Advanced Science Research Center, Japan Atomic Energy Agency, 2-4 Shirakata, Tokai 319-1195, Japan.*

[6]*RIKEN Center for Emergent Matter Science (CEMS), Wako 351-0198, Japan.*

[7]*Cavendish Laboratory, Department of Physics, University of Cambridge, Cambridge, CB3 0HE, United Kingdom.*

[8]*Beijing National Laboratory for Condensed Matter Physics, Institute of Physics, University of Chinese Academy of Sciences, Chinese Academy of Sciences, Beijing 100190, China.*

[9]*Shenzhen Institute for Quantum Science and Engineering, Southern University of Science and Technology, Shenzhen 518055, China.*

[10]*Institute of Physics, Ecole Polytechnique Fédérale de Lausanne (EPFL), 1015, Lausanne, Switzerland.*

[11]*Kavli Institute for Theoretical Sciences (KITS), University of Chinese Academy of Sciences, Beijing 100190, China.*

[12]*Institute of Electrical and Micro Engineering (IEM), École Polytechnique Fédérale de Lausanne (EPFL), Lausanne, Switzerland.*

†These authors contributed equally.
*dirk.grundler@epfl.ch
 haiming.yu@buaa.edu.cn

**Controlling spin current lies at the heart of spintronics and its applications. The sign of spin currents is monotonous in ferromagnets once the current direction is determined. Spin currents in antiferromagnets can possess opposite polarization, but requires enormous magnetic fields to lift the degeneracy. Controlling spin currents with different polarization is urgently demanded but remains hitherto elusive. Here, we demonstrate the control of spin currents at room temperature by magnon interference in a canted antiferromagnet, hematite recently also classified as an altermagnet. Magneto-optical characterization by Brillouin light scattering revealed that the spatial periodicity of the beating patterns was tunable via the microwave frequency. The inverse spin-Hall voltage changed sign as the frequency was scanned, i.e., a frequency-controlled switching of polarization in pure spin currents was obtained. Our work marks the use of antiferromagnetic magnon interference to control spin currents, which substantially extends the horizon for the emerging field of coherent antiferromagnetic spintronics.**

Spin waves, or their quanta magnons[1,2], can coherently transfer spin information without Joule heating. They are therefore promising as information carriers for next-generation spintronic devices with ultra-low power consumption. Spin-wave interference[3-7], as a key signature for magnon coherence, is at the heart of numerous coherent spintronic devices, such as spin-wave logic circuits[8,9], magnonic directional couplers[10] and neural networks[11]. Recently, coherent antiferromagnetic spintronics[12-14] has attracted extensive interest for exploiting spin coherence in antiferromagnets for ultrafast information processing with high reliability. The authors of Ref. 15 studied optical excitation of spin wave pulses in the single-crystal antiferromagnet $DyFeO_3$ below 90 K and suggested about coherence-mediated logic devices. To reach this goal, experimental evidence of magnon interference is a key prerequisite. However, magnon interference in a true antiferromagnet has not been observed, despite some studies on antiferromagnetically coupled ferromagnetic multilayers[16,17]. In this work, we report the direct observation at room temperature of interference between coherently excited antiferromagnetic magnons in single-crystal $\alpha$-Fe$_2$O$_3$, also known as hematite[18-21]. Hematite is a canted antiferromagnet at room temperature and also classified recently as an altermagnet[22-24]. The interference pattern occurs over several wavelengths and is spatially imaged by micro-focused Brillouin light scattering spectroscopy ($\mu$-BLS). We observe that the spatial beating pattern periodicity becomes smaller at a higher excitation frequency. By placing a Pt bar at a certain position ($s$) from the magnon emitter (Fig. 1**a**), we detect pure spin current signals and investigate the inverse spin Hall effect (ISHE) which oscillates as a function of frequency between positive and negative values. These signals reflect the alternating spin polarization induced by the interference of antiferromagnetic magnons. We ascribe these independent measurements to interferences between an in-plane polarized bulk spin-wave mode[25-27] and an out-of-plane polarized surface spin-wave mode[28]. A distinct nonreciprocal behavior is observed in both the optical ($\mu$-BLS) and electrical (ISHE) measurements, which supports the surface magnon interpretation. Our findings offer a means of controlling spin-precessional polarization without changing magnetic field or temperature, and thus provide new opportunities for coherent antiferromagnetic spintronics with enhanced versatility of magnons as information carriers for next-generation logic and computation.

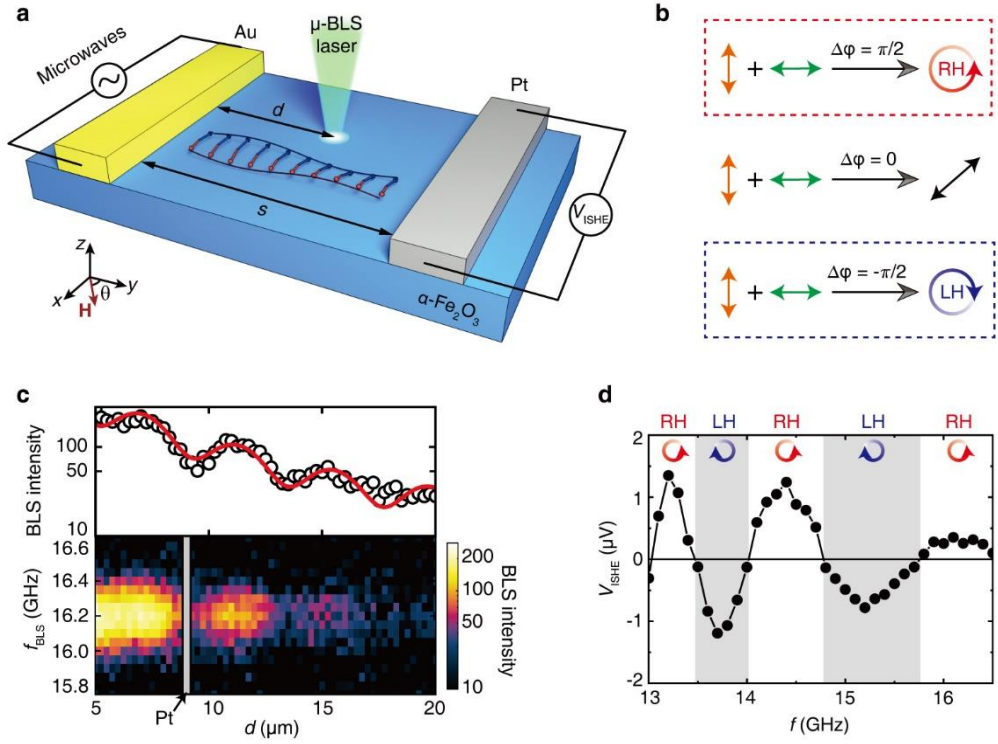

**Fig. 1 | Antiferromagnetic spin-wave interference in spatial and frequency domains. a**, Schematic diagram of the device studied with micro-focused Brillouin light scattering (µ-BLS) and inverse spin-Hall voltage ($V_{\text{ISHE}}$) detection. Antiferromagnetic spin waves are coherently excited in hematite (α-$Fe_2O_3$) by a gold antenna (200 nm wide, 100 µm long and 120 nm thick). The $xy$ plane is the magnetic anisotropy easy plane. A Pt bar (300 nm wide, 100 µm long and 7 nm thick) is used as a detector for pure spin currents via the inverse spin-Hall effect. The BLS laser detects magnon intensities at a distance ($d$) from the antenna and with a spatial resolution of 250 nm. In this study, a magnetic field is applied in-plane at an angle θ = 45° with respect to the $y$ axis. **b**, Illustration of the interference between two magnon modes with vertical and horizontal spin-precessional polarizations. After a propagation distance $s$, a phase difference of $\Delta\varphi = \Delta k\, s$ builds up and produces right-handed (RH) total spin precession when $\Delta\varphi = \pi/2$ and left-handed (LH) one when $\Delta\varphi = -\pi/2$. **c**, The BLS intensity measured as a function of the distance. The excitation frequency is set at 16.2 GHz and a magnetic field of -50 mT is applied. **d**, $V_{\text{ISHE}}$ measured by the Pt detector at $s = 9$ µm as a function of excitation frequency. Here, the applied magnetic field is -20 mT. The input microwave power is set at 15 dBm. White and gray regions mark respectively the right-handed (RH) and left-handed (LH) precessions with reversed $V_{\text{ISHE}}$ signs.

We study a single-crystal hematite film with the $c$ axis [0001] pointing perpendicular to the film plane (Methods). Below the Morin temperature (~266 K), the sample is in the easy-axis phase with no detectable net magnetic moment (Supplementary Fig. S1). At room temperature, however, the sample is in an easy-plane phase exhibiting a small canted moment which allows for a low-frequency antiferromagnetic magnon mode[20,21]. We employ an experimental scheme (Fig. 1**a**) with broadband excitation using a nano-stripline (NSL) antenna[29] (Supplementary Fig. S2) and characterize the magnon propagation by the µ-BLS. A Pt bar is placed at a distance ($s$) from the NSL for the magnon detection by the ISHE[30,31] (Methods). The ISHE is sensitive to pure spin currents[32] and its sign depends on the magnon handedness[33,34] (Fig. 1**b**), i.e., left-handed and right-handed total spin precessions. Figure 1**c** shows the spatial variation of colour-coded BLS spectra probed by the focused laser spot when scanning from distance $d = 5$ µm to $d = 20$ µm for an excitation frequency of 16.2 GHz in a field of -50 mT. The lineplot is extracted at the same detection frequency of 16.2 GHz. An oscillatory signal variation is clearly observed on the background of an exponential decay of the magnon amplitude as a function of distance $d$. We consider the oscillation to arise from a standing beating pattern as a consequence of the interference between two antiferromagnetic (AFM) spin waves. If these two modes show different magnon polarizations (predominantly vertical and horizontal) and wavelength ($\lambda$), the superposition of these two coherently excited modes will not only exhibit the beating pattern of the magnon intensity but also lead to the oscillation of magnon handedness as illustrated in Fig. 1**b**. This interpretation is corroborated by our ISHE measurements. Figure 1**d** presents $V_{\mathrm{ISHE}}$ measured on a Pt detector[30-32] placed at $s = 9$ µm and plotted as a function of the frequency from 13 GHz to 17 GHz with an applied magnetic field of -20 mT. The $V_{\mathrm{ISHE}}$ sign reversal indicates the alternation between right-handed (RH) and left-handed (LH) precession orientations. The positive $V_{\mathrm{ISHE}}$ indicates right-handed precession whereas the negative $V_{\mathrm{ISHE}}$ is the signature of left-handed ones. The handedness is defined in reference to the control experiment performed on an yttrium iron garnet (YIG) film which is known to present only right-handed precession[30,31] (Supplementary Fig. S3). The BLS measurement at a set frequency (16.2 GHz) gives direct evidence of the magnon interference in the real space. The oscillatory ISHE voltage gives direct evidence of the interference by scanning the frequency at a set detector position ($s = 9$ μm).

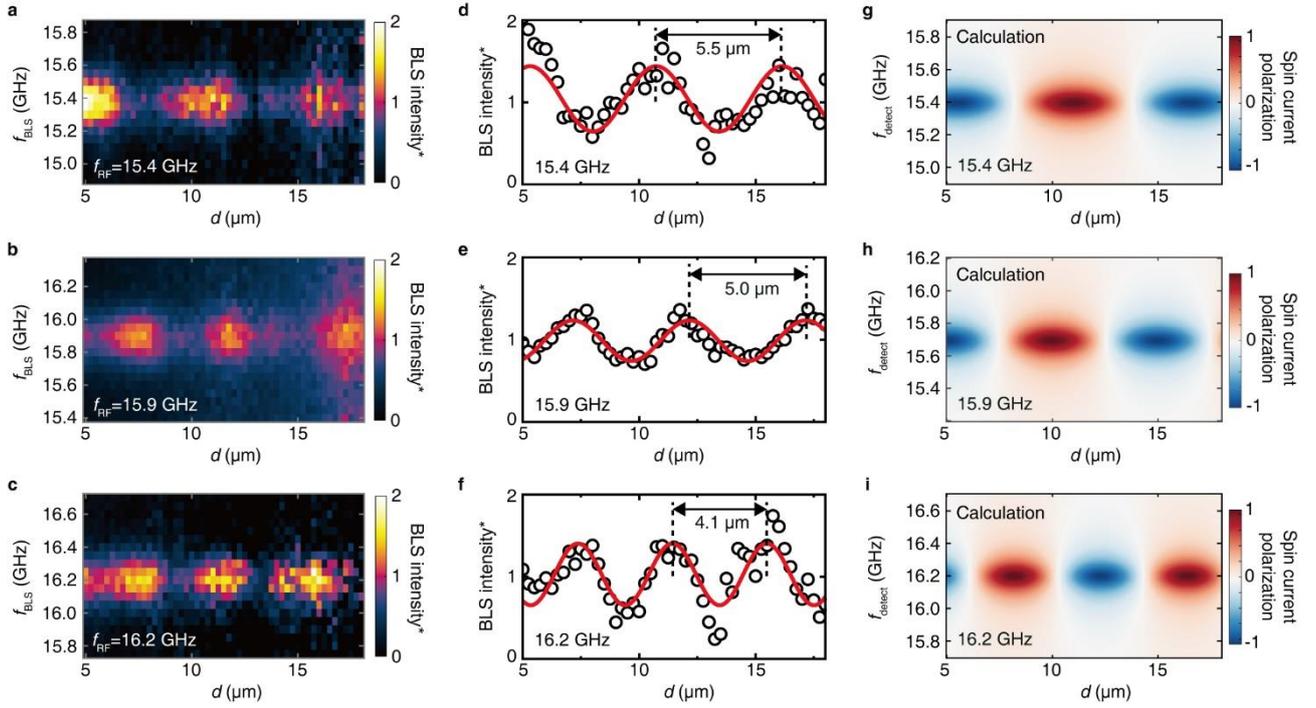

**Fig. 2 | BLS spatial mapping of the AFM magnon interference at different frequencies. a-c**, Spatially resolved BLS measurement of magnon interference with continuous-wave microwave excitation at frequencies ($f_{RF}$) of 15.4 GHz (**a**), 15.9 GHz (**b**) and 16.2 GHz (**c**). A magnetic field of -50 mT is applied at 45° with respect to the propagation direction. The BLS detection frequency ($f_{BLS}$) is swept about the excitation frequency. The BLS intensity is presented for the distance ($d$) from 5 μm to 18 μm in a colour scale after considering an exponential decay compensation. **d-f**, Lineplots extracted with the condition of $f_{BLS} = f_{RF}$ for 15.4 GHz (**d**), 15.9 GHz (**e**) and 16.2 GHz (**f**). The periodicity of the interference pattern is fitted to be 5.5 (±0.2) μm, 5.1 (±0.2) μm and 4.1 (±0.3) μm, respectively. **g-i**, Calculated spatial distribution of the normalized spin current polarization at difference excitation frequencies. The blue colour indicates a positive spin polarization, whereas the red colour indicates a negative spin polarization.

When the microwave excitation frequency ($f_{RF}$) is tuned, the magnon interference pattern observed in the BLS spatial mapping changes accordingly as shown in Fig. 2**a-c** for 15.4 GHz, 15.9 GHz and 16.2 GHz, respectively. The colour-scale BLS intensity is compensated[35] for the distance-dependent decay of the magnon intensity (Methods). For the case when the BLS detection frequency equals the microwave excitation frequency ($f_{BLS} = f_{RF}$), we extract line plots from the compensated BLS intensity as a function of the probing distance ($d$) from the antenna and present the data in Figs. 2**d-f** at different frequencies. The periodicity of the oscillatory pattern is extracted to be 5.5 ($\pm$0.2) μm, 5.1 ($\pm$0.2) μm and 4.1 ($\pm$0.3) μm for 15.4 GHz, 15.9 GHz and 16.2 GHz, respectively. This excitation frequency dependence of BLS spatial mapping is consistent with that of ISHE measurement showing alternating voltage signals when the Pt detector is placed at a fixed position (Fig. 1**d**). According to a simple theoretical model of interference between bulk acoustic magnon mode of hematite and another vertically polarized spin oscillation (Methods), the beating of BLS intensity should be accompanied by alternating spin current polarization[36,37] with the same spatial periodicity as shown in Figs. 2**g-i**. The frequency dependence of periodicity can be predicted from the bulk magnon dispersion relation, which agrees qualitatively with the observation. A full chart of spin current polarization calculated as a function of both frequency and distance is presented in Supplementary Fig. S4**i**, which agrees reasonably well with both the distance-dependent BLS measurement and the frequency-dependent spin pumping measurement.

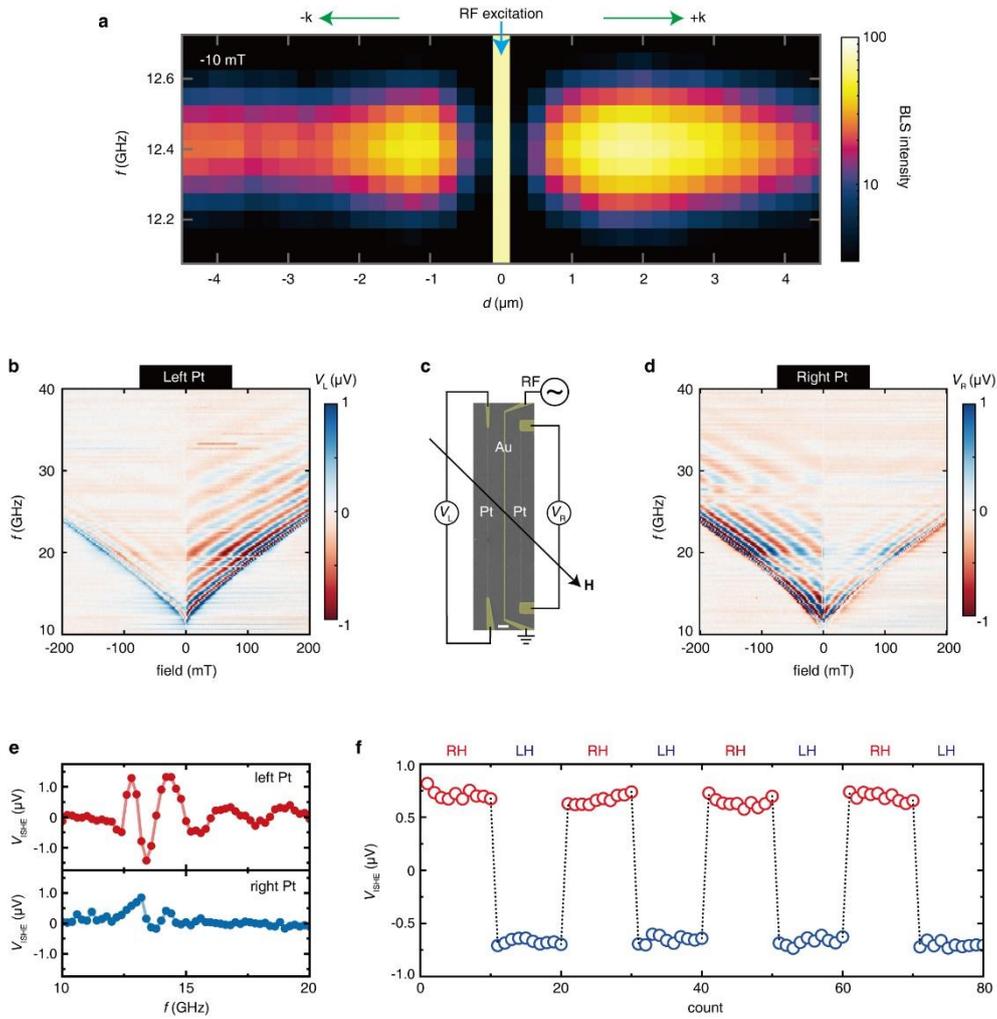

**Fig. 3 | Nonreciprocal behavior of AFM magnon interference. a**, The BLS spatial mapping measured at -10 mT along 45° with the RF antenna placed at the centre position ($d = 0$ μm). The excitation microwave frequency is set at 12.4 GHz. **b**, Inverse spin-Hall voltages measured on the left Pt detector ($V_L$) plotted as a function of magnetic field and excitation frequency. The distance between the antenna and Pt bar is 8 μm. **c**, Coloured SEM image of the device with Pt detectors on both sides of the NSL antenna. White scale bar, 5 μm. Magnetic field (**H**) is swept at 45° with respect to the antenna. **d**, Inverse spin-Hall detection of spin currents at the right Pt bar ($V_R$). **e**, Single line plots extracted from the data measured on the left Pt detector (upper panel) and the right Pt detector (bottom panel). The magnetic field is set at -10 mT. **f,** Manipulation of spin currents by tuning the excitation frequency between 14.8 GHz (right-handed) and 14.3 GHz (left-handed). The microwave power is set at 15 dBm. A magnetic field of -40 mT is applied at θ = 45°.

To gain more insights into the nature of the interference, we placed the microwave antenna at the centre of the hematite sample, and conducted the BLS spatial mapping on both sides of the antenna. Figure 3**a** shows that the magnon intensities on the right side is stronger than those on the left side reflecting nonreciprocal spin-wave propagation. The strong nonreciprocity is also electrically detected with two Pt bars (Figs. 3**b** and 3**d**) placed on the left and right sides of the antenna as shown in the microscope image in Fig. 3**c**. Figures 3**b** and 3**d** present the inverse spin-Hall voltages measured on the left and right Pt detectors, respectively and plotted as a function of excitation frequency and applied magnetic field. A number of oscillations of $V_{ISHE}$ are observed over broad frequency and field regimes. More oscillations of $V_{ISHE}$ are detected in the frequency domain with an increasing propagation distance (*s*) as shown in Supplementary Fig. S4. The angular dependence of $V_{ISHE}$ (Supplementary Fig. S5) is measured and qualitatively consistent with the conventional $(A\sin^2\theta + B)\cos\theta$ shape that takes account of the efficiencies of antenna excitation and ISHE detection[30,31]. Oscillations of $V_{ISHE}$ persist over virtually all directions, however, with a small in-plane magnetic anisotropy and a clear nonreciprocity[27]. In most of the measurements, we employ the $\theta = 45°$ configuration to attain efficiency in both excitation and detection. At a negative applied field, e.g., -10 mT, the ISHE voltages detected on the right Pt bar are significantly larger than those detected on the left side (Fig. 3**e**), whereas at positive fields, the left side exhibits stronger ISHE voltages than the right side. The asymmetric magnon propagation (intensity) to the left and to the right with wave vectors $(-k)$ and $(+k)$, respectively, indicates that at least one of the interfering magnon modes is chiral, which could potentially be a surface mode of the Damon-Eshbach type[38] or a surface magnon polariton[39] in an antiferromagnet. A recent report on nonreciprocal propagation of spin waves in hematite attributed its findings to the presence of dipolar-exchange surface spin waves[27]. We are unable to fully confirm this identification, however, as the surface mode exists in a very narrow range of $\theta$ according to the existing theoretical model[40] while our nonreciprocal interference pattern is detected across all angles except for $\theta \approx 0°, \pm 90°, \pm 180°$. The bulk magnon mode has an elliptic precession of a large eccentricity with its major axis lying in-plane. Hence, the interfering magnon mode must have a predominantly out-of-plane linear polarization in order to account for the alternating spin current, something which the tiny dipolar field alone cannot induce. Although the optical mode of the antiferromagnetic resonance does possess an appropriate polarization[36], its frequency is well above 100 GHz and too high to be relevant in our experiment. Another possible candidate may be a classical analogue of spin-singlet edge states in 1D antiferromagnetic spin chains[41], although such microscopic modelling of hematite is beyond the scope of this study and left for a future work.

Our assignment of the observed BLS and ISHE effects to interference are further confirmed with the following experiments. First, power-dependent data showed that the $V_{\text{ISHE}}$ oscillation existed in the linear excitation regime (Supplementary Fig. S6). This observation further supports our hypothesis that linear interference causes alternating spin currents. Second, we found that the use of an NSL was necessary to obtain this interference effect. We conducted conventional spin pumping in a flip-chip configuration (Supplementary Fig. S7) using a 75-µm-wide CPW antenna which was two orders of magnitude wider than the NSL and prevented the emission of short-wave magnons. The results on hematite were consistent with previous studies[20,21], where only the ordinary right-handed spin precession was observed. Third, we also found that below the Morin temperature, where the canted moment vanishes, the oscillatory $V_{\text{ISHE}}$ signal disappeared (Supplementary Fig. S8). This experimental observation confirms that the alternating spin currents at room temperature arise primarily from the precession of the canted moment associated with propagating AFM magnons coherently excited at the NSL.

Finally, we show that potential spurious effects which could appear as an interference can be eliminated. In order to check whether the crosstalk-induced ac current in Pt caused a rectification owing to the sizable spin-Hall magnetoresistance (SMR) of hematite[42], we fabricated devices with different ISHE detectors (Supplementary Fig. S9), namely Pt and W. At a given field and frequency, the $V_{ISHE}$ sign was reversed for Pt and W due to their opposite spin-Hall angles[43,44]. Since SMR is known to be proportional to the square of spin-Hall angle, this eliminates the crosstalk-induced current rectification as a primary mechanism of the interference. Other possible ac signals in hematite include phonons, which nevertheless couple very weakly with magnons. The phonons are not expected to be excited efficiently in those experiments as the hematite crystal respects inversion symmetry and consequently there is no piezoelectric effect. Even if they were excited efficiently through acoustic resonance, they would be unable to affect the spin-wave precession across such a wide range of frequency and magnetic field.

To conclude, we provide an experimental demonstration (Fig. 3**f**) of the control over pure spin current by simply tuning the excitation microwave frequency, which leads deterministically to negative and positive ISHE voltages at a given applied field. This frequency-controlled spin injection could be used to switch the magnetization of a nanomagnet located at the surface of hematite, thus providing a convenient technique to switch the nanomagnet[45] back and forth with frequency-tunable magnon spin torque[46,47] (see Supplementary Fig. S10 for a proposed device). The antiferromagnetic magnon interference observed in this work offers a means of controlling spin current polarization without changing magnetic field or temperature, and thus expands the horizons for magnonics enhancing the versatility of magnons as information carriers for next-generation logic and computing[48].

**References**

1   Kruglyak, V. V., Demokrotiv, S. O. & Grundler, D. Magnonics. *J. Phys. D: Appl. Phys.* **43**, 264001 (2010).
2   Pirro, P., Vasyuchka, V. I., Serga, A. A. & Hillebrands, B.  Advances in coherent magnonics. *Nat. Rev. Mater.* **6**, 1114 (2021).
3   Podbielski, J., Giesen, F. & Grundler, D. Spin-wave interference in microscopic rings. *Phys. Rev. Lett.* **96**, 167207 (2006).

4    Perzlmaier, K., Woltersdorf, G. & Back, C. H. Observation of the propagation and interference of spin waves in ferromagnetic thin films. *Phys. Rev. B* **77**, 054425 (2008).

5    Demidov, V. E., Demokritov, S. O., Rott, K., Krzysteczko, P. & Reiss, G. Mode interference and periodic self-focusing of spin waves in permalloy microstripes. *Phys. Rev. B* **77**, 064406 (2008).

6    Bertelli, I. et al. Magnetic resonance imaging of spin-wave transport and interference in a magnetic insulator. *Sci. Adv.* **6**, eabd3556 (2020).

7    Chen, J. et al. Reconfigurable spin-wave interferometer at the nanoscale. *Nano Lett.* **21**, 6237-6244 (2021).

8    Khitun, A., Bao, M. & Wang, K. L. Magnonic logic circuits. *J. Phys. D: Appl. Phys.* **43**, 264005 (2010).

9    Talmelli, G. et al. Reconfigurable submicrometer spin-wave majority gate with electrical transducers. *Sci. Adv.* **6**, eabb4042 (2020).

10   Wang, Q. et al. A magnonic directional coupler for integrated magnonic half-adders. *Nat. Electron.* **3**, 765-774 (2020).

11   Papp, Á., Porod, W. & Csaba, G. Nanoscale neural network using non-linear spin-wave interference. *Nat. Commun.* **12**, 6422 (2021).

12   Han, J., Cheng, R., Liu, L., Ohno, H. & Fukami, S. Coherent antiferromagnetic spintronics. *Nat. Mater.* **22**, 684-695 (2023).

13   Li, J. et al. Spin current from sub-terahertz-generated antiferromagnetic magnons. *Nature* **578**, 70-74 (2020).

14   Vaidya, P. et al. Subterahertz spin pumping from an insulating antiferromagnet. *Science* **368**, 160-165 (2020).

15   Hortensius, J. R. et al. Coherent spin-wave transport in an antiferromagnet. *Nat. Phys.* **17**, 1001-1016 (2021).

16   Albisetti, E. et al. Optically inspired nanomagnonics with nonreciprocal spin waves in synthetic antiferromagnets. *Adv. Mater.* **32**, 1906439 (2020).

17   Girardi, D. et al. Three-dimensional spin-wave dynamics, localization and interference in a synthetic antiferromagnet. *Nat. Commun.* **15**, 3057 (2024).

18   Lebrun, R. et al. Tunable long-distance spin transport in a crystalline antiferromagnetic iron oxide. *Nature* **561**, 222-225 (2018).

19   Jani, H. et al. Antiferromagnetic half-skyrmions and bimerons at room temperature. *Nature* **590**, 74-79 (2021).

20   Wang, H. et al. Spin pumping of an easy-plane antiferromagnet enhanced by Dzyaloshinskii-Moriya interaction, *Phys. Rev. Lett.* **127**, 117202 (2021).

21   Boventer, I. et al. Room-temperature antiferromagnetic resonance and inverse spin-Hall voltage in canted antiferromagnets, *Phys. Rev. Lett.* **126**, 187201 (2021).

22   Šmejkal, L., Sinova, J. & Jungwirth, T. Emerging research landscape of altermagnetism, *Phys. Rev. X* **12**, 040501 (2022).

23   Lee, R. A., Afanasiev, D., Kimel, A. V. & Mikhaylovskiy, R. V. Canted spin order as a platform for ultrafast conversion of magnons. *Nature* https://doi.org/10.1038/s41586-024-07448-3.

24   Galindez-Ruales, E. F. et al. Altermagnetism in the hopping regime. arXiv:2310.16907.

25   Hamdi, M., Posva, F. & Grundler, D. Spin wave dispersion of ultra-low damping hematite ($\alpha$-Fe$_2$O$_3$) at GHz frequencies. *Phys. Rev. Mater.* **7**, 054407 (2023).

26   Wang, H. et al. Long-distance coherent propagation of high-velocity antiferromagnetic spin waves. *Phys. Rev. Lett.* **130**, 096701 (2023).

27   El Kanj, A. et al. Evidence of non-degenerated, non-reciprocal and ultra-fast spin-waves in the canted antiferromagnet α-Fe$_2$O$_3$. *Sci. Adv.* **9**, eadh1601 (2023).

28  Camley, R. E. Long-wavelength surface spin waves on antiferromagnets. *Phys. Rev. Lett.* **45**, 283-286 (1980).

29  Ciubotaru, F., Devolder, T., Manfrini, M., Adelmann, C. & Radu, I. P. All electrical propagating spin wave spectroscopy with broadband wavevector capability. *Appl. Phys. Lett.* **109**, 012403 (2016).

30  d'Allivy Kelly, O. et al. Inverse spin Hall effect in nanometer-thick yttrium iron garnet/Pt system. *Appl. Phys. Lett.* **103**, 082408 (2013).

31  Wang, J. et al. Broad-wave-vector spin pumping of flat-band magnons. *Phys. Rev. Appl.* **21**, 044024 (2024).

32  Saitoh, E. et al. Conversion of spin current into charge current at room temperature: Inverse spin-Hall effect. *Appl. Phys. Lett.* **88**, 182509 (2006).

33  Cheng, R., Xiao, J., Niu, Q. & Brataas, A. Spin pumping and spin-transfer torques in antiferromagnets. *Phys. Rev. Lett.* **113**, 057601 (2014).

34  Liu, Y. et al. Switching magnon chirality in artificial ferrimagnet. *Nat. Commun.* **13**, 1264 (2022).

35  Demidov, V. E. et al. Excitation of coherent propagating spin waves by pure spin currents. *Nat. Commun.* **7**, 10446 (2016).

36  Han, J. et al. Birefringence-like spin transport via linearly polarized antiferromagnetic magnons. *Nat. Nanotechnol.* **15**, 563-568 (2020).

37  Wimmer, T. et al. Observation of antiferromagnetic magnon pseudospin dynamics and the Hanle effect. *Phys. Rev. Lett.* **125**, 247204 (2020).

38  Damon R. W. & Eshbach J. R. Magnetostatic modes of a ferromagnet slab. *J. Phys. Chem. Solids* **19**, 308 (1961).

39  Macêdo, R. & Camley, R. E. Engineering terahertz surface magnon-polaritons in hyperbolic antiferromagnets. *Phys. Rev. B* **99**, 014437 (2019).

40  Tarasenko, V. V. & Kharitonov, V. D. *JETP* **33**, 1246-1250 (1971).

41  White, S. R. Density-matrix algorithms for quantum renormalization groups. *Phys. Rev. B* **48**, 10345 (1993).

42  Fischer, J. et al. Large spin Hall magnetoresistance in antiferromagnetic α-Fe$_2$O$_3$/Pt heterostructures. *Phys. Rev. Appl.* **13**, 014019 (2020).

43  Liu, L. et al. Spin-torque switching with the giant spin Hall effect of tantalum. *Science* **336**, 555-558 (2012).

44  Wang, H. et al. Scaling of spin Hall angle in 3d, 4d, and 5d metals from Y$_3$Fe$_5$O$_{12}$/metal spin pumping. *Phys. Rev. Lett.* **112**, 197201 (2014).

45  Baumgärtl, K. & Grundler, D. Reversal of nanomagnets by propagating magnons in ferrimagnetic yttrium iron garnet enabling nonvolatile magnon memory. *Nat. Commun.* **14**, 1490 (2023).

46  Han, J., Zhang, P., Hou, J. T., Siddiqui, S. A. & Liu, L. Mutual control of coherent spin waves and magnetic domain walls in a magnonic device. *Science* **366**, 1121-1125 (2019).

47  Wang, Y. et al. Magnetization switching by magnon-mediated spin torque through an antiferromagnetic insulator. *Science* **366**, 1125-1128 (2019).

48  Chumak, A. V. et al. Advances in magnetics roadmap on spin-wave computing. *IEEE Trans. Magn.* **58**, 0800172 (2022).

## Methods

**Sample information and device fabrication.** The *c*-cut α-Fe$_2$O$_3$ single-crystalline bulk sample is a commercial product from the company SurfaceNet, the dimension of this sample is 5 mm × 5 mm × 0.5 mm. The Morin transition temperature of this sample is around 265 K from the SQUID magnetometry. Thus, this sample exhibits an easy-plane phase at room temperature (~293 K) and the *c* axis [0001] is along the out-of-plane direction. At the easy-plane phase, the two sublattices of α-Fe$_2$O$_3$ have a small titled angle due to Dzyaloshinskii-Moriya interaction. When the temperature is lower than Morin temperature, the Néel vector rotates from in-plane to out-of-plane [0001] crystalline orientation then α-Fe$_2$O$_3$ becomes a uniaxial antiferromagnet. On top of the α-Fe$_2$O$_3$ sample, we use electron beam lithography to fabricate nanostructures including nano-stripline (NSL), coplanar waveguide (CPW) microwave antennas, and nano $V_{\text{ISHE}}$ detectors. Then we grow 120 nm thick Au for microwave antennas and 7 nm thick Pt or 3 nm thick W for the $V_{\text{ISHE}}$ detection. The Pt and W films were deposited by a magnetron sputtering system with a base pressure of $3.0 \times 10^{-6}$ Pa at room temperature. The sputtering power was kept at 100 W and the deposition rate was 0.1 nm s$^{-1}$. The Au was deposited at 30 W, and the deposition rate was 0.5 nm s$^{-1}$.

**Nonlocal spin pumping measurements.** The nonlocal spin pumping device consists of two parts: the NSL (CPW) microwave antenna and the Pt (W) bar for $V_{\text{ISHE}}$ detection. We have patterned a series of devices with different propagation distances (3 μm~11 μm) between the microwave antennas and Pt (W) bars. The width of the NSL (CPW) and Pt (W) bars are all 300 nm. The SEM image of the representative device can be found in Fig. S2**a**. The microwave antennas are connected to a PSG analog signal generator (Keysight E8257D) via a three-port microwave probe. For the DC detection part, a nanovoltmeter (Keysight 34420A) is used to pick up the DC voltages induced by spin pumping in α-Fe$_2$O$_3$ and YIG. The output frequency (power is set as 15 dBm for all measurements) of the signal generator is fixed and the in-plane magnetic field is swept from negative to positive values. For each magnetic field and frequency, the DC voltage measured by the nanovoltmeter is recorded. Then, the $V_{\text{ISHE}}$ versus field line plot is plotted as shown in Fig. 1**d**. By combining multiple line plots at different frequencies, one attains the colour-scale plots as shown in Figs. 3**b** and 3**d**. The background noise can be suppressed by subtracting the DC voltages measured with the microwave on by signals measured with the microwave off.

**Inelastic light scattering microscopy on microwave-induced spin waves.** The spatial distribution of magnon resonance signals was measured by a scanning micro-focused Brillouin Light Scattering (BLS) setup at room temperature. A monochromatic continuous-wave solid-state laser with a wavelength of 532 nm and a power of 1 mW was focused with a lens with a numerical aperture of 0.85 onto a diffraction-limited spot. Backscattered light was analyzed with a six-pass Fabry-Perot interferometer TFP-2 (JRS Scientific Instruments). The BLS signal was assumed to be proportional to the square of the amplitude of the dynamic component of the out-of-plane magnetization at the position of the laser spot. The sample was mounted on a closed-loop piezo-stage. Spatial maps were obtained by positioning the laser spot on the sample surface at different distances from the stripline antenna. The step size between different positions amounted of 250 nm. The positioning system was stabilized by a custom-designed active feedback, providing long-term spatial stability. Permanent magnets were used to apply external magnetic fields in the plane of the hematite sample at 45° with respect to the long axis of the stripline antenna. The ends of the antenna were wire-bonded to a printed-circuit board, which was electrically connected to a signal generator applying a microwave current. The microwave current produced a radiofrequency magnetic field which exerted a torque on the spins in hematite and induced spin-wave emission. For displaying colour-coded BLS spectra reflecting the spatial distribution of local minima in the spin-wave signals we proceeded as follows. Distance-dependent BLS data were taken for a fixed irradiation frequency on a line perpendicular to the antenna. We fitted the function $A \exp(-\gamma d) + y_0$ to the maxima of the spin-wave spectra, where $\gamma$ was the inverse of the decay length and $d$ was the distance from the stripline antenna. Using the parameters extracted from the fit, we subtracted $y_0$ from the original data and multiplied the remaining difference by the factor $\exp(\gamma d)/A$. The result was analysed in terms of a standing wave pattern in which we had compensated for the signal decay (damping) as a function of distance $d$ from the spin-wave emitting antenna. Thereby we enhanced the signal-to-noise ratio of local minima in the distance-dependent BLS signals.

**Calculation for spatial distribution of magnon handedness.** Here, we present a toy model that exhibits the observed spin-wave spatial interference pattern. We use the Landau-Lifshitz equations of easy-plane antiferromagnets with the Dzyaloshinskii-Moirya interaction appropriate for the crystalline symmetry of hematite[21]

$$\frac{d\mathbf{m}_1}{dt} = -\gamma\mu_0 \mathbf{m}_1 \times \left[\mathbf{H}_0 - H_{\text{ex}}\mathbf{m}_2 - \frac{1}{2}H_{\text{ex}}a_{\text{ex}}^2\nabla^2\mathbf{m}_2 - H_A(\mathbf{m}_1 \cdot \hat{\mathbf{z}})\hat{\mathbf{z}} + H_a(\mathbf{m}_1 \cdot \hat{\mathbf{y}})\hat{\mathbf{y}} + H_{\text{DM}}(\mathbf{m}_2 \times \hat{\mathbf{z}})\right], \quad (1)$$

$$\frac{d\mathbf{m}_2}{dt} = -\gamma\mu_0 \mathbf{m}_2 \times \left[\mathbf{H}_0 - H_{\text{ex}}\mathbf{m}_1 - \frac{1}{2}H_{\text{ex}}a_{\text{ex}}^2\nabla^2\mathbf{m}_1 - H_A(\mathbf{m}_2 \cdot \hat{\mathbf{z}})\hat{\mathbf{z}} + H_a(\mathbf{m}_2 \cdot \hat{\mathbf{y}})\hat{\mathbf{y}} - H_{\text{DM}}(\mathbf{m}_2 \times \hat{\mathbf{z}})\right], \quad (2)$$

where $\mathbf{m}_1$ and $\mathbf{m}_2$ are the normalized sublattice magnetizations, $\gamma$ is the gyromagnetic ratio, $\mu_0$ is the vacuum permeability, the coordinate system is such that $x$ and $z$ are aligned to the external magnetic field and surface normal respectively, and the parameters are taken as follows; external magnetic field $\mu_0|\mathbf{H}_0| = 50$ mT, exchange field $H_{\text{ex}} = 1040$ T, inter-sublattice exchange stiffness length $a_{\text{ex}} = 0.3$ nm, out-of-plane anisotropy field $H_A = 1$ mT, in-plane anisotropy field $H_a = 6.6 \times 10^{-5}$ T, and Dzyaloshinskii-Moriya field $H_{\text{DM}} = 2.7$ T.

We can further deduce the lower-frequency eigenmode (acoustic mode) of $\mathbf{m}_{1,2}$ by the ansatz $\mathbf{m}_{1,2}(k, \omega) = \mathbf{m}_{10,20} \exp i(\mathbf{k} \cdot \mathbf{r} - \omega t)$. For detailed methodology and geometry to simplify Eqs.1 and 2 to an eigenvalue problem, one may refer to Ref. 21. In the coordinate system in which $\hat{\mathbf{x}}$ and $\hat{\mathbf{z}}$ are aligned to the external magnetic field and surface normal respectively, one may take $\mathbf{m}_{10,20} = \left(1, m_{10,20}^y, m_{10,20}^z\right)$. The precession of the total magnetization could be then deduced from

$$\begin{cases} m_y(k, \omega) = \sin\theta \left(m_{10}^y + m_{20}^y\right), \\ m_z(k, \omega) = m_{10}^z + m_{20}^z, \end{cases} \quad (3)$$

where $\theta \ll 1$ is the canting angle of hematite. We assume there is another magnon mode, which is vertically polarized (i.e., vanishing $x, y$ components), and propagates instantaneously (i.e., no phase delay). The amplitude of this mode is denoted by $m_z^{\text{rf}}$. Then, the spatial distribution of the total magnetization could be represented by

$$\begin{cases} \tilde{m}_y(r) = m_y(k, \omega) \cdot \exp i\mathbf{k} \cdot \mathbf{r}, \\ \tilde{m}_z(r) = m_z(k, \omega) \cdot \exp i\mathbf{k} \cdot \mathbf{r} + m_z^{\text{rf}}, \end{cases} \quad (4)$$

where the tilde denotes the total magnetization taking into account the interference with the hypothetical second mode. According to the standard spin pumping theory[33], the spin current should be given by

$$I_s \approx G\omega \frac{\tilde{m}_y^*\tilde{m}_z - \tilde{m}_z^*\tilde{m}_y}{2i} = G\omega\left(\text{Im}[m_y(k,\omega)^* m_z(k,\omega)] - |m_y(k,\omega)^* m_z^{\text{rf}}|\sin(\mathbf{k} \cdot \mathbf{r} + \alpha)\right), \quad (5)$$

where $G$ is the real part of spin-mixing conductance and $\alpha$ is the phase of $m_y(k, \omega)^* m_z^{\text{rf}}$. The second term represents the beating and consequently spin current is spatially modulated. Taking into account the almost in-plane polarization of the bulk acoustic magnon $|m_z(k,\omega)| \ll |m_y(k,\omega)|$, $I_s$ can be dominated by the beating term and spatially oscillate with period $2\pi/|\mathbf{k}|$. In Figs. 2**g-i**, total spin current after propagation $\tilde{I}_s = \Gamma(\omega - \omega_0)I_s$ is plotted as a function of propagation length $d$ and detected frequency $\omega$, and we take $|m_z^{\text{rf}}|/|m_y(\omega_0, k_0)| = 1$ and $\alpha = 0$, where $\omega_0$ and $k_0$ are excited frequency and the corresponding wavevector. Given the $m_{y,z}$ are normalized magnetization, the absolute amplitude of $m_z^{\text{rf}}$ is meaningless while the ratio between $|m_z^{\text{rf}}|$ and $|m_z(\omega_0, k_0)|$ is crucial. The excitation wavevector $k_0$ is calculated by the equation $\omega(k_0) = \omega_0$. The Lorentzian envelop function,

$$\Gamma(\omega - \omega_0) = \frac{1}{(\omega - \omega_0)^2 + \Delta\omega^2}, \qquad (6)$$

has been inserted to qualitatively account for the resonant nature of the excitation without overcomplicating the modelling of the unknown $m_z^{\text{rf}}$, with $\Delta\omega = 0.1$ GHz is the linewidth.

## Data availability

All data are available in the main text or the Supplementary Information. Other data relevant to this paper are available from the corresponding author upon reasonable request.

## Acknowledgement

The authors thank J. Wang, J. Xiao, W. Jiang, I. Boventer, O. Gomonay, J. Li and P. Yan for their helpful discussions. We wish to acknowledge the support by the National Key Research and Development Program of China, Grants No. 2022YFA1402801; NSF China under Grants No. 12074026, No. 52225106; China Scholarship Council (CSC) under Grant No. 202206020091; the Shenzhen Institute for Quantum Science and Engineering, Southern University of Science and Technology (Grant No. SIQSE202007); JST PRESTO Grant No. JPMJPR20LB; JST CREST Grants (No. JPMJCR19J4, No. JPMJCR1874, and No. JPMJCR20C1); and JSPS KAKENHI (Nos. 21K13886, 20H10865). D.G. and A.D. thank SNSF for funding via grant 197360.

**Author contributions**

L.S., H.W. and H.Y. conceived and designed the experiments. L.S., H.W., J.W., S.L., D.Y. and H.Y. fabricated the nanostructured devices on hematite. A.D. and D.G. conducted and analysed the BLS measurements. L.S., H.W., K.-L.Y., J.H. and H.Y. performed nonlocal spin pumping measurements. L.S., Y.Z. and H.Y. conducted the time-gating measurements. H.W. measured the conventional spin pumping. P.C., W.H. and X.H. provided the YIG control sample. L.S., A.D., H.W., R.Y., J.C., J.-Ph.A., D.G. and H.Y. analysed the data. K.Y. and S.M. provided the theoretical discussion. D.G. and H.Y. supervised the experimental study. H.Y., L.S., A.D., K.Y., J.-Ph.A., S.M. and D.G. wrote the paper and discussed with all authors.

**Competing financial interests**

The authors declare no competing financial interests.